\documentstyle[12pt,epsfig]{article}

\begin{document}

\begin{center}
{\huge Can only flavor-nonsinglet $H$ dibaryons\\}
{\huge be stable against strong decays?\\}
\vspace*{1.5cm}
{\large Stathes D. Paganis\footnote{e-mail address: 
paganis@physics.utexas.edu}, 
Takeshi Udagawa\footnote{e-mail address: 
udagawa@physics.utexas.edu},\\ 
G. W. Hoffmann\footnote{e-mail address: 
hoffmann@physics.utexas.edu}, 
and R. L. Ray\footnote{e-mail address: 
ray@physics.utexas.edu} \\
         Department of Physics, University of Texas at Austin\\
         Austin, Texas 78712}\\

\vspace*{0.5cm}
{\large Manuscript Prepared for Brief Reports of Physical Review C}
\end{center}
\vspace*{2cm}

\begin{abstract}
Using the QCD sum rule approach,
we show that the flavor-nonsinglet $H$ dibaryon states with
J$^{\pi} = 1^+$, J$^{\pi} = 0^+$, I=1 (27plet) 
are nearly degenerate with the J$^{\pi} = 0^+$, I=0 singlet 
$H_0$ dibaryon, 
which has been predicted to be stable against strong decay, 
but has not
been observed.  Our
calculation, which does not require an instanton correction, suggests
that the $H_0$ is slightly heavier than these flavor-nonsinglet $H$s
over a wide range of the
parameter space.  
If the singlet $H_0$ mass lies above the $\Lambda \Lambda$ threshold
(2231~MeV), then the strong interaction breakup to $\Lambda \Lambda$
would produce a very broad resonance in the $\Lambda \Lambda$ 
invariant mass
spectrum which would be very difficult to observe.  
On the other hand, if these
flavor-nonsinglet J=0 and 1 $H$ dibaryons are also above the 
$\Lambda \Lambda$ threshold,
but below the $\Xi^0n$ breakup threshold (2254 MeV), then because
the direct, strong interaction decay to the $\Lambda \Lambda$ 
channel is
forbidden, these flavor-nonsinglet states might be more amenable to
experimental observation.
The present results allow a possible reconciliation 
between the reported
observation of
$\Lambda \Lambda$ hypernuclei, which argue against a stable $H_0$,
and the possible existence of $H$ 
dibaryons in general.

\end{abstract}

\vspace{0.2in}
\noindent
PACS number(s): 12.40.Yx, 14.20.Pt

\newpage
\section{Introduction}
\label{intro}
\vspace{0.2in}

Using the MIT bag model, Jaffe\cite{jaffe}
predicted a stable (against strong decay) 
six-quark flavor-singlet
(uuddss) hadron (referred to as $H_0$) with J=0, I=0, and S=$-2$.
A plethora of mass calculations \cite{mass} followed 
Jaffe's work, and most of them 
predicted a weakly bound $H_0$ of mass 
just below the $\Lambda \Lambda$ threshold
(2231 MeV). For 20 years experiments have searched for the
$H_0$, but no convincing evidence has been
found for its existence \cite{Hsearch,aoki}.
In fact, the candidate $H_0$s from different experiments have 
very different masses. 
One such experiment \cite{aoki} claims the observation 
of a very weakly bound $\Lambda \Lambda$ hypernucleus which
excludes at some level 
the existence of the $H_0$. 

SU(3) flavor-nonsinglet 
$H$ states are not usually discussed in 
the literature because of the 
expectation that they should be heavier than the $H_0$. 
This expectation is due to the assumption that
the effective magnetic one-gluon-exchange 
between the valence quarks
is most attractive for the flavor-singlet channel, 
making the $H_0$ the lightest state. 

In this work we consider the masses of two
nonsinglet, doubly strange six-quark states which occur in the
baryon-octet$\otimes$baryon-octet
direct product space \cite{Xie}.  These include the
J$^{\pi} = 1^+$, I=0 $H$
from the J=1 $SU(3)_f$ octet, and
the I=1, I$_3$=0 $H$ from the J$^{\pi} = 0^+$ 27plet.
Based on model-independent assumptions, we show 
that it is likely that the $H_0$ 
is nearly degenerate with these states, and the $H_0$ mass 
is slightly {\it larger}.
This contradiction with the MIT bag model has been discussed 
elsewhere \cite{oka}.
We study the mass ratios, rather than absolute masses, because the 
ratio
carries smaller uncertainty and is more stable over a wide region 
of the parameter space.
The QCD sum rule approach was used by Ioffe\cite{ioffe}
to calculate the mass splitting in the baryonic decuplet.

The situation for these $H$ states is analogous to the case of the 
$\eta^{\prime}$ (I=0, J=0), 
$\rho$ (J=1, I=1) and $\omega$ (J=1, I=0), in which
the $\eta^{\prime}$ is the heaviest. 
The extra mass is attributed to the $U(1)_A$ 
anomalous symmetry breaking of QCD, which is taken 
into account in model 
calculations using 
instantons. The method employed here does not need an 
instanton correction because 
as discussed in \cite{derek} the 
instanton effects (if any) are effectively included 
in the quark condensates.

If the singlet $H_0$ mass is greater 
than the $\Lambda \Lambda$ mass,
then it should not have been 
observed in previous $H$-search experiments \cite{Hsearch,aoki}.
If the nonsinglet
J=1, I=0 (octet) $H$ mass is above
the $\Lambda \Lambda$ threshold
(2231 MeV), but below the $\Xi^0n$ threshold (2254 MeV), then it
can only decay weakly or electromagnetically
because the strong decay to $\Lambda \Lambda$ is not allowed.
The situation is
different for the J=0, I=1 (27plet) $H$ since it 
may isospin mix with the $H_0$
and the J=0, I=0 (27plet) $H$ to form 
physical states which strong decay
to $\Lambda \Lambda$.  However, for this case, 
if the mixing is small,
experimental
evidence for these $H$s may be a narrow peak 
in the $\Lambda \Lambda$
invariant mass spectrum.
Such an observation would not contradict the 
observed $\Lambda \Lambda$
hypernucleus events \cite{aoki}.  
The candidate nonsinglet J=0, I=1 $H$s
reported by Shabazian {\it et al.} \cite{Hsearch} 
might be explained in this
way.

In Sec.~\ref{sumrule}
we discuss the QCD sum rule method and then formulate sum rules for 
the $H_0$, the J=1, I=0 (octet) $H$ and the J=0, I=1 (27plet) $H$. 
In Sec.~\ref{calcs} we calculate the mass ratios $m_{H_0}/m_H$ over
a wide range of the parameter space.
Finally, in Sec.~\ref{results} 
we discuss our results and some experimental
issues and present our conclusions.

\section{{\it H} Dibaryon Sum Rule }
\label{sumrule}
\vspace{0.2in}

In this section, the QCD sum rules \cite{SVZ} are 
formulated for the $H_0$,
the  J=1, I=0 (octet) $H$ and the  J=0, I=1, I$_3$=0 (27plet) $H$. 
We follow the method described in Ref.~\cite{oka}.
The $H_0$ current, $J_{H_0}(x)$,
is
a product of two baryonic currents, $J_{B}(x)$, which are written
using the convention in 
\cite{esp} as
\begin{equation}
J_B(x)=O_1^{q_iq_jq_k}(x) + tO_2^{q_iq_jq_k}(x),
\label{eqn : os}
\end{equation}
where,
\begin{equation}
O_1^{q_iq_jq_k}(x)
= \epsilon_{abc} \cdot
[ (q^{Ta}_{i}(x)Cq^{b}_{j}(x)) \gamma_{5}
q^{c}_{k}(x)],
\label{eqn : bacur1}
\end{equation}
and
\begin{equation}
O_2^{q_iq_jq_k}(x)=\epsilon_{abc} \cdot 
[(q^{Ta}_{i}(x)C\gamma_{5}q^{b}_{j}(x))q^{c}_{k}(x)].
\label{eqn : bacur2}
\end{equation}
In Eqs.~(\ref{eqn : bacur1}) and (\ref{eqn : bacur2}) 
$q$ is the quark field, $i,j,k$ 
are flavor indices, $a,b,c$ are color 
indices, $C$ is the charge conjugation matrix, 
and the parameter $t$ is a weighting coefficient 
for the second term of the baryonic current which is commonly 
used in QCD sum rule calculations.
The $H_0$ dibaryon singlet current is given by
\begin{eqnarray}
J_{H_0}(x) &=& (2 \epsilon_{ijk'}
 \epsilon_{i'j'k} - \frac{2}{3} \epsilon_{ijk}
 \epsilon_{i'j'k'}) \cdot J_BC\gamma_5J_B 
\label{eqn : hcurr}
\end{eqnarray}
where the antisymmetrization tensors, $\epsilon_{ijk}$, produce
a color and flavor-singlet 6-quark 
(uuddss) state with the quantum numbers of the $H_0$. 

The current correlation function is written as
\begin{equation}
\Pi_{H}(q^2) \equiv -i \int d^{4}x e^{iqx} \langle 0| T\ J_{H}(x)\ 
\overline{J}_{H}(0) |0 \rangle. 
\label{eqn : contr}
\end{equation}
We apply the operator product expansion (OPE) 
to Eq.~(\ref{eqn : contr})
to obtain the right hand side (rhs) of the $H$ dibaryon sum rule
for large $Q^2=-q^2$. 
The advantage of the OPE for dibaryons is that there are just 
three non-vanishing terms that give 
large contributions in the $SU(3)_f$ limit.
The resulting correlation function in the $SU(3)_{f}$ 
limit is \cite{oka}
\begin{eqnarray}
\Pi_{H}(q^2)&=& \frac{h_1(t)}{2^{14}\pi^{10}\Gamma(9)\Gamma(8)}
(-q^{2})^{7}ln(-q^{2})
+\frac{h_2(t)}{2^{8}\pi^{6}\Gamma(6)\Gamma(5)}
(-q^{2})^{4}ln(-q^{2}) \cdot \frac{\langle\overline{q}q\rangle^{2}}
{(4N_{c})^{2}} \nonumber \\
&&+\frac{h_3(t)}{2^{2}\pi^{2}
\Gamma(3)\Gamma(2)}(-q^{2})ln(-q^{2})
\cdot \frac{\langle\overline{q}q\rangle^{4}}{(4N_{c})^{4}},
\label{eqn : hcorr}
\end{eqnarray}
where $N_c$ is the number of color charges and  
$\langle\overline{q}q\rangle$ is the quark condensate.
The $h_i(t)$ coefficients in Eq.~(\ref{eqn : hcorr}) 
are obtained by calculating all the possible 
contractions in 
Eq.~(\ref{eqn : contr}). For example, for the $h_1(t)$ and $t=0$,
there are 3600 non-vanishing 
terms.
The $h_i(t)$s for the $H_0$ have 
been calculated elsewhere \cite{oka}.
 
Using the group symmetry properties of the 
dibaryon octet$\otimes$octet
direct product space \cite{Xie} the
current for the J=0, I=1, I$_3$=0 (27plet) $H$ is written as
\begin{eqnarray}
J_{H^{27}, I=1, I_3=0}(x)&=&2J_{\Lambda} C \gamma^{5} J_{\Sigma^0} 
+2J_{\Sigma^0} C \gamma^{5} J_{\Lambda} \nonumber \\
&&+J_P C \gamma^{5} J_{\Xi^-}
+J_{\Xi^-} C \gamma^{5} J_{P} \nonumber \\
&&-J_{N} C \gamma^{5} J_{\Xi^0}
-J_{\Xi^0} C \gamma^{5} J_{N},
\end{eqnarray}
where for the baryon currents we use the 
convention of Eq.~(\ref{eqn : os}).
For example, \\
 $P~=~(q_iq_jq_k)~=~(udu)$ for the proton.
The calculation yields the following $h_i$ coefficients:
\begin{eqnarray}
h_1(t)&=&1302 + 120 t - 804 t^2 + 984 t^3 + 858 t^4 \nonumber \\
h_2(t)&=&5208 + 480 t - 3216 t^2 + 3936 t^3 - 6408 t^4 \nonumber \\
h_3(t)&=&-20832 - 1920 t + 12864 t^2 - 15744 t^3 + 25632 t^4. 
\end{eqnarray}
For the J$^{\pi} = 1^+$, I=0 (octet) $H$ the current is written as
\begin{eqnarray}
J_{H^*}^{\mu}(x)&=&J_{\Sigma^0} C \gamma^{\mu} J_{\Sigma^0}
- \frac{1}{2}J_{\Sigma^+} C \gamma^{\mu} J_{\Sigma^-} 
- \frac{1}{2}J_{\Sigma^-} C \gamma^{\mu} J_{\Sigma^+} \nonumber \\
&&- \frac{1}{3}J_{\Lambda} C \gamma^{\mu} J_{\Lambda} 
+ \frac{1}{4}J_{N} C \gamma^{\mu} J_{\Xi^0} 
+ \frac{1}{4}J_{\Xi^0} C \gamma^{\mu} J_{N} \nonumber \\
&&+ \frac{1}{4}J_{P} C \gamma^{\mu} J_{\Xi^-} 
+ \frac{1}{4}J_{\Xi^-} C \gamma^{\mu} J_{P}.
\label{eqn : h1curr}
\end{eqnarray}
This state, called the $H^*$, does not strongly couple to 
the $\Lambda \Lambda$ channel \cite{bicker,errata}.
For this case the $h_i$ coefficients were determined to be:
\begin{eqnarray}
h_1(t)&=&1866 + 2087 t - 365 t^2 + 1418 t^3 + 2636 t^4 \nonumber \\
h_2(t)&=&7466 + 8349 t - 1460 t^2 + 5671 t^3 - 20026 t^4 \nonumber \\
h_3(t)&=&-29864 - 33397 t + 5840 t^2 - 22683 t^3 + 80104 t^4. 
\end{eqnarray}

\newpage
\section{Calculation of the Mass Ratios}
\label{calcs}
\vspace{0.2in}

The $H_0$ mass is given\cite{oka} by the expression

\begin{eqnarray}
m_{H}^{2} (M^2)
&=& \left[\frac{h_1(t)}{2^{14}\pi^{10}\Gamma(8)}
(M^{2})^{9} (1-\Sigma_{8}) -
\frac{h_2(t)}{2^{8}\pi^{6}\Gamma(5)}
\cdot \frac{\langle\overline{q}q\rangle^{2}}{(4N_{c})^{2}}
(M^{2})^{6} (1-\Sigma_{5}) \right. \nonumber \\
&& \left. + \frac{h_3(t)}{2^{2}\pi^{2}\Gamma(2)}
\cdot \frac{\langle\overline{q}q\rangle^{4}}{(4N_{c})^{4}}
(M^{2})^3(1-\Sigma_{2}) \right] / \nonumber \\
&& \left[ \frac{h_1(t)}{2^{14}\pi^{10}\Gamma(9)}
(M^{2})^{8} (1-\Sigma_{7}) -
\frac{h_2(t)}{2^{8}\pi^{6}\Gamma(6)}
\cdot \frac{\langle\overline{q}q\rangle^{2}}{(4N_{c})^{2}}
(M^{2})^{5} (1-\Sigma_{4}) \right. \nonumber \\
&& \left. + \frac{h_3(t)}{2^{2}\pi^{2}\Gamma(3)}
\cdot \frac{\langle\overline{q}q\rangle^{4}}{(4N_{c})^{4}}
(M^{2})^{2}(1-\Sigma_{1}) \right] ,
\label{eqn : hmass}
\end{eqnarray}
where
\[ \Sigma_i =\mathop{\sum}_{k=0}^{i}
\frac{s_{0}^{k}}{(M^{2})^{k}k!}e^{-s_{0}/M^{2}}, \]
accounts for the continuum part, 
$M$ is the Borel mass, and  $s_0$ 
is the continuum threshold.\footnote{The 
third power in the 
$M^2$ term in Eq.~(\ref{eqn : hmass}) is missing in 
Eqs.~(10) and (15) of Ref.~\cite{oka} due to
a typographical error.}
Eq.~(\ref{eqn : hmass}) also holds for the other $H$s, so 
the mass ratios, given by
\begin{eqnarray}
R&=&
\frac{m_{H_0}(t,M_0^{(1)},s_0^{(1)},\langle\overline{q}q\rangle^{2})} 
{m_{H,J}(t,M_0^{(2)},s_0^{(2)},\langle\overline{q}q\rangle^{2})} ~,
\label{eqn : ratio}
\end{eqnarray}
were calculated for J=0 and J=1.

We use the standard assumptions for the phenomenological side (lhs) 
of Eq.~(\ref{eqn : contr}) (pole term plus continuum for the
spectral density, with
continuum threshold $s_0$).
The central values for our parameters are $t~=~-1.2$, 
$s_0$=~5.694~GeV$^2$, $M$=~1.5~GeV,
$\langle\overline{q}q\rangle^{2}=~(-0.250)^3~$GeV$^3$.
We expect the calculation
to be reliable for a wide range of the Borel mass M
because the $H$ mass is determined by the chiral symmetry 
breaking scale ($\sim~1~GeV$)\cite{manohar} which is much larger than the 
scale at which
QCD vacuum fluctuations become large ($\Lambda_{QCD}\simeq 200$~MeV)
and where the Borel smearing fails as it does for the
case of the light pseudoscalar mesons. 
This region is taken to be around 
2~GeV, where
the higher order terms in the OPE are strongly 
suppressed and the pole 
dominates the continuum contribution.
Our choice for the parameter $t$, as discussed 
in \cite{derek}, gives 
self-consistent QCD sum rules which do not need 
an instanton correction.
Instead, the instanton effects are adequately accounted for in the 
nonperturbative vacuum condensates.
The Borel mass $M$ is in general different for the two $H$ states,
but we expect 
it to be the same for degenerate states with the same quark content.

\section{Results and Discussion}
\label{results}
\vspace{0.2in}

\begin{figure}
\psfig{figure=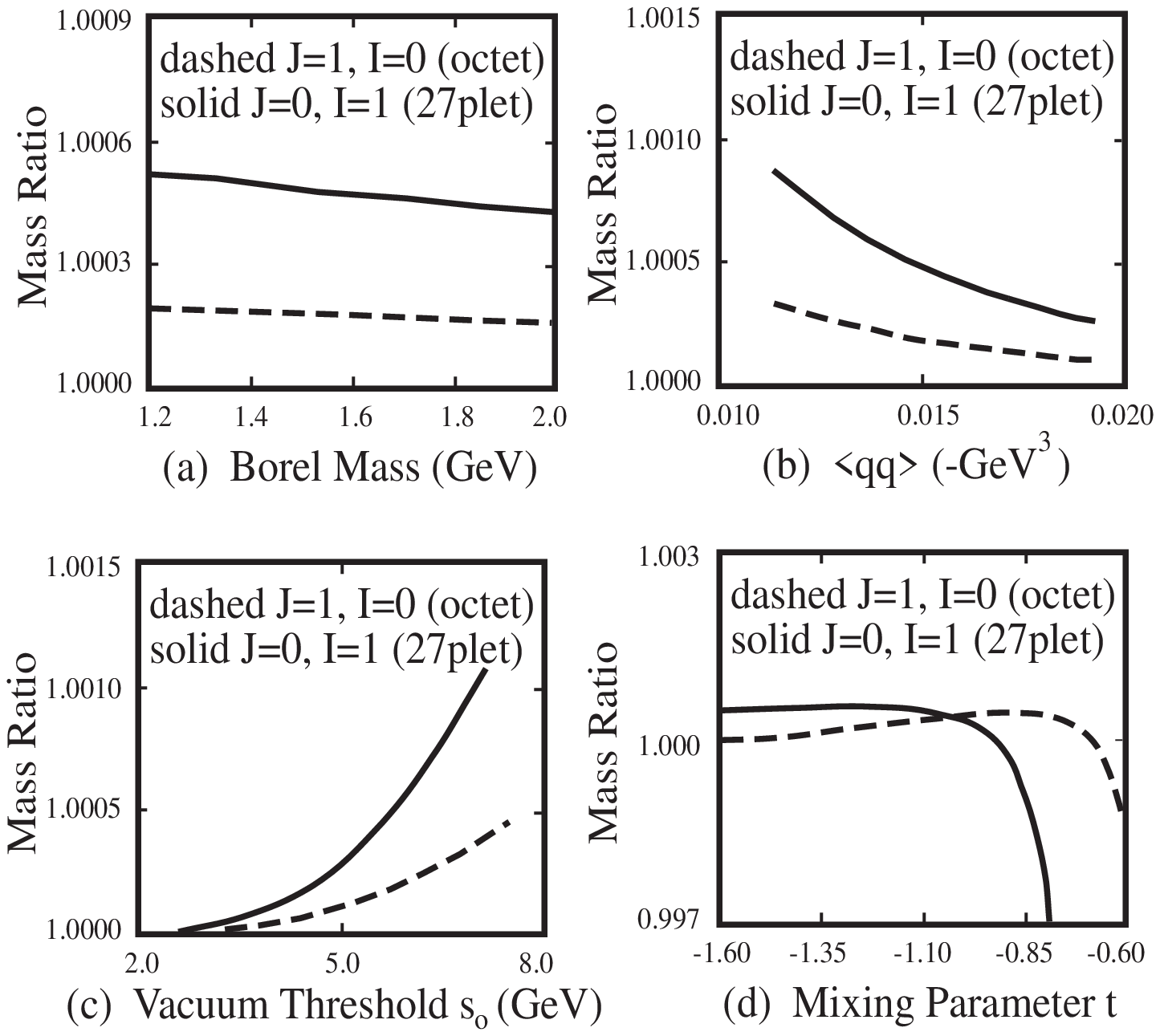}
\caption{
Dependence of the mass ratio [Eq. (\protect{\ref{eqn : ratio}})] 
on (a) the Borel
mass, (b) the quark condensate, (c) the vacuum threshold, and (d)
the mixing parameter.
The solid curve is $m_{H_0}/m_{H,J=0~(27plet)}$, while
the dashed curve is $m_{H_0}/m_{H,J=1~(octet)}$.
}
\label{fig1}
\end{figure}

Our results for the mass ratios 
from Eq.~(\protect{\ref{eqn : ratio}})  
and their sensitivities to variations of the input
parameters are summarized in 
Fig.~\ref{fig1}.
Fig.~\ref{fig1}a shows the dependence of the 
mass ratios 
on the Borel mass. The solid curve is $m_{H_0}/m_{H,J=0~(27plet)}$, 
while
the dashed curve is $m_{H_0}/m_{H,J=1~(octet)}$. 
The ratios remain constant,
and are slightly 
greater than 1, for a large range of the Borel mass. Fig.~\ref{fig1}b shows
the dependence of the mass ratios 
on the quark condensate.  
The curves have the same meaning as in Fig.~\ref{fig1}a.
Fig.~\ref{fig1}b shows that large variation of the
quark condensate, within 
a range consistent with that found in the literature, 
produces very little effect ($\leq 2$ MeV for the mass difference).
Fig.~\ref{fig1}c and Fig.~\ref{fig1}d
show the mass ratio dependence on the continuum 
threshold $s_0$ and the parameter $t$.
Again we see little sensitivity, about  $0.1\%$.
Finally, the 
sensitivity of the mass ratio with respect 
to different Borel masses for the
$H_0$ and the J=0, I=1 (27plet) $H$ 
or the J=1, I=0 (octet) $H$  was studied. The Borel mass of the 
$H_0$ was fixed and the nonsinglet $H$ Borel mass was varied.
These mass ratios were essentially the same at each Borel mass and
they decreased linearly from 1.05 for a Borel mass of 1.3 GeV to 0.99
for a Borel mass of 1.6 GeV.
Other contributors
to the theoretical uncertainty are 
the neglected terms in the OPE 
(which we expect to 
be small) and the vacuum saturation assumption: 
$\langle\overline{q}^{2}q^{2}\rangle 
\sim \langle\overline{q}q\rangle^2$, 
$\langle\overline{q}^{4}q^{4}\rangle 
\sim \langle\overline{q}q\rangle^4$.

Our conclusion, based on the results 
presented in Fig.~\ref{fig1}
is that the mass ratio remains very close 
to unity in the parameter range where 
we believe the calculation to be
reliable. 
Both the J=1, I=0 (octet) $H$  and the J=0, I=1 (27plet) $H$ 
are almost degenerate with the $H_0$, and this degeneracy is 
not sensitive to the 
various QCD sum rule parameters.
The mass 
difference is of the order of $0.1\%$ for both $H$ states, or 
about 2~MeV for $m_H$=~2~GeV.
{\it Surprisingly}, 
we find that the singlet $H_0$ is slightly heavier than both
the J=0, I=1 (27plet) $H$
and the J=1, I=0 (octet) $H$.

It is worthwhile to consider the consequences 
of these predictions for
experiments designed to search for neutral, strange 
dibaryons.  We note that
the nonsinglet J$^{\pi} = 1^+$, I=0 octet $H$ cannot 
strong decay to
$\Lambda \Lambda$ (due to angular momentum and parity 
conservation), and if
its mass is lower than the $\Xi^0$n threshold (2254~MeV), then 
it may only
decay electromagnetically (via an M1 transition to 
the J$^{\pi} = 0^+$, I=1
27plet $H$ and/or the singlet $H_0$ if these are 
lower in mass, or to the
$\Lambda \Lambda \gamma$ channel) or weakly (if 
its mass is below that of the
other $H$s and the $\Lambda \Lambda$ threshold).  
If the J$^{\pi} = 0^+$, I=1
27plet $H$ has mass greater than 2231~MeV it cannot 
strong decay to
$\Lambda \Lambda$ due to isospin conservation.  
However, isospin mixing,
due to electromagnetic interactions within the 
dibaryon,  cause the physical
J$^{\pi} = 0^+$ 27plet $H$ to contain a small 
I=0 admixture, which allows
strong decay to $\Lambda \Lambda$.  If the 
admixture is small, then the 
state could have a narrow width of perhaps 
a few MeV.  The 27plet $H$
cannot electromagnetically decay directly to 
the singlet $H_0$ (except via
two-photon decay), however an 
E1 transition to the $\Lambda \Lambda \gamma$ 
is permitted.  If the J=0 27plet 
$H$ is below the other $H$s and the
$\Lambda \Lambda$ threshold, then 
it may only decay by the weak interaction.

\begin{figure}
\psfig{figure=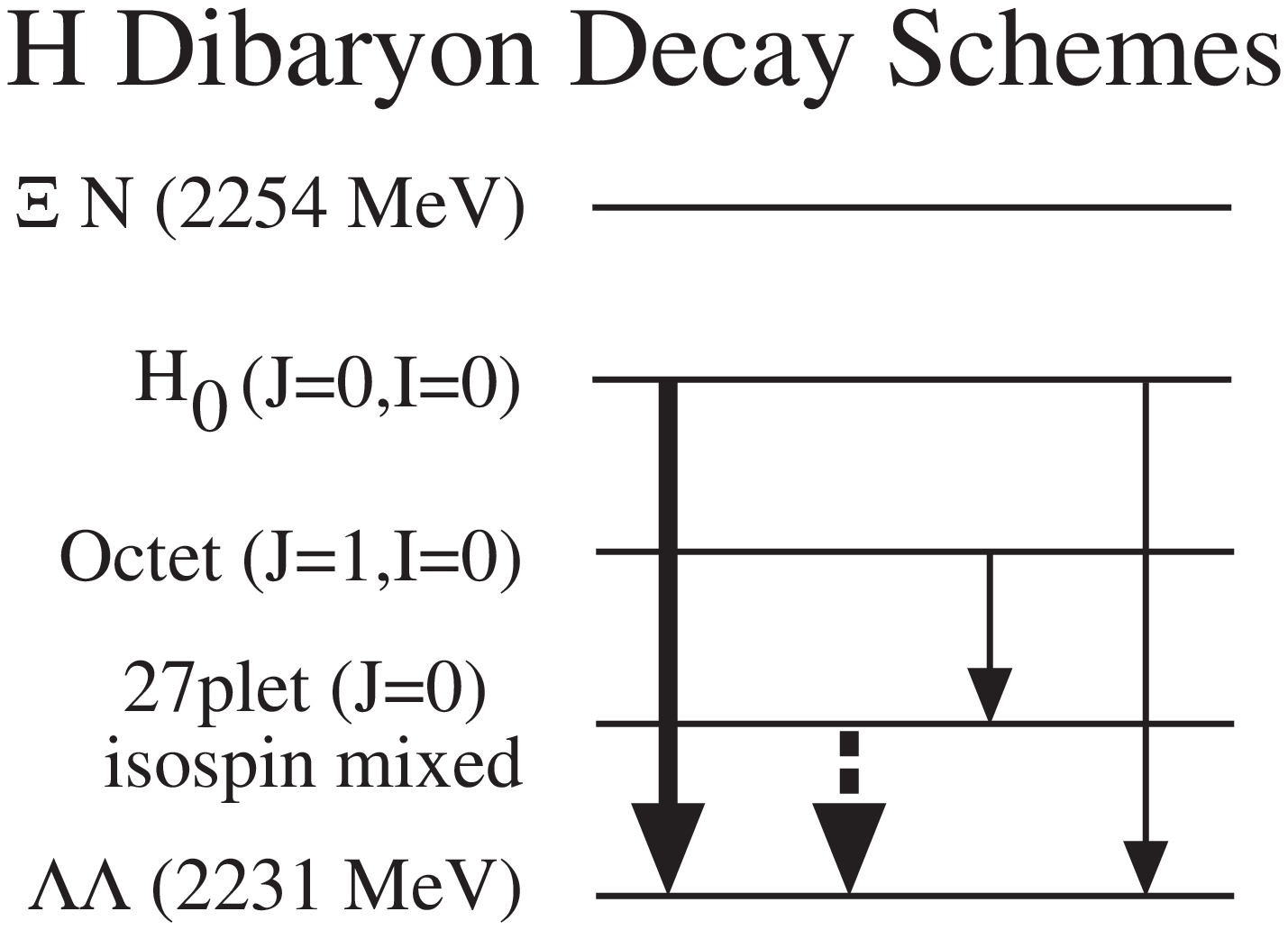}
\caption{Strong and electromagnetic
decay schemes for the $H_0$, the J=1, I=0 octet $H$, and the
J=0 isospin mixed 27plet $H$ assuming the relative masses predicted
here and assuming the $H$ masses lie between 2231 and 2254 MeV, as discussed
in the text.  Thick, solid arrow indicates strong decay; thick, dashed
arrow indicates strong decay via small I=0 isospin admixture; and thin
arrows indicate electromagnetic decays.
}
\label{fig2}
\end{figure}

If we assume the predicted mass ratios found here and that
each mass is between the $\Lambda \Lambda$ mass and the
$\Xi^0$n mass, then the strong and 
electromagnetic decay schemes for the
three $H$s will be as shown in Fig.~\ref{fig2}.
If the decay
of the J$^{\pi} = 0^+$ 27plet $H$ 
is driven by the strong decay channel
via the small isospin mixing I=0 component rather than by 
the electromagnetic
decay to the $\Lambda \Lambda \gamma$ channel, 
then a relatively narrow
peak should appear in the $\Lambda \Lambda$ 
invariant mass spectrum between
2231 and 2254~MeV.  The decay of the J$^{\pi} = 1^+$, 
I=0 octet $H$ to the
$\Lambda \Lambda \gamma$ channel would result in a 
broad three-body phase
space distribution for the $\Lambda \Lambda$ mass 
spectrum if the photon is
not observed. The electromagnetic decays are as follows: (1) 
J$^{\pi} = 1^+$, I=0 octet $H$ to the J$^{\pi} = 0^+$ 27plet 
$H$ is a
$\Delta$I=0 and 1, M1 transition,
(2) J$^{\pi} = 1^+$, I=0 octet $H$ to 
$\Lambda \Lambda \gamma$ occurs via
E1 and M1 transitions
(all are $\Delta$I=0), and
(3) J$^{\pi} = 0^+$ 27plet $H$ 
to $\Lambda \Lambda \gamma$ occurs via an E1
transition.
In Fig.~\ref{fig2} the strong decay of the $H_0$ is
indicated by the thick,
solid arrow, the strong decay 
via the small isospin mixing component is
shown by the thick, dashed arrow, and the electromagnetic decays
by the thin, solid arrows.

Observation of these nonsinglet $H$s 
would not contradict the $\Lambda \Lambda$
hypernucleus events already observed \cite{aoki}.  
It is possible that the
candidate nonsinglet $H$ (J=0, I=1) observed by 
Shabazian {\it et al.}
\cite{Hsearch} can be explained by these results 
since we predict the
J$^{\pi} = 0^+$ 27plet $H$ to be approximately 
degenerate with $H_0$.
However, the mass must be less than that 
reported in \cite{Hsearch}
in order to be below the $\Xi^0$n
breakup threshold.  Clearly the discovery 
of flavor-nonsinglet $H$ dibaryons
would require us to revisit traditional 
hadronic structure models and
require a better understanding of quark-quark 
effective interactions.

\vspace{0.1in}

The authors thank Dr. 
Ron Longacre at the Brookhaven National Laboratory
for helpful discussions.
This work was supported in part by the U.S. Department of 
Energy and
The Robert A. Welch Foundation.

\end{document}